\documentclass[12pt]{article}
\usepackage{amscd,amsmath,amssymb,amstext,amsthm,exscale,latexsym}
\usepackage{graphicx}
\newcommand {\al}   {\alpha}       \newcommand {\bt}  {\beta}
\newcommand {\g }   {\gamma}       
\newcommand {\dl}   {\delta}

         \newcommand {\ph}  {\phi}
\newcommand {\vf }  {\varphi}

\newcommand {\pl}   {\partial}

\newcommand {\MR}  {{\mathbb R}}
   
\newcommand {\Sm}  {{\textsc{m}}}   \newcommand {\Sn}  {{\textsc{n}}}

\theoremstyle{definition}

\begin{document}
\title     {Complete separation of variables in the geodesic Hamilton--Jacobi
            equation in four dimensions}
\author    {M. O. Katanaev
            \thanks{E-mail: katanaev@mi-ras.ru}\\ \\
            \sl Steklov Mathematical Institute,\\
            \sl 119991, Moscow, ul. Gubkina, 8}
\maketitle
\begin{abstract}
We list all metrics of arbitrary signature in four dimensions which admit
complete separation of variables in the Hamilton--Jacobi equation for geodesic
Hamiltonians. There are only ten classes of separable metrics admitting
commuting Killing vector fields, indecomposable quadratic conservation laws, and
coisotropic coordinates. Canonical separable metrics parameterized by several
(up to twelve) arbitrary functions of single coordinates are written explicitly.
The full set of independent conservation laws in involution for each canonical
metrics is also found.
\end{abstract}

%*******************************************************************************
In 1891, St\"ackel raised the question: ``Which metrics admit complete
separation of variables in the respective Hamilton--Jacobi equation?''
\cite{Stacke91,Stacke93A}, and gave the answer in the case of diagonal
metrics for quadratic conservation laws. These metrics are called St\"ackel or
separable.
The problem attracted much interest of mathematicians and physicists and
became classical. It is clear that if the metric has enough symmetry then it may
admit $n$ envolutive conservation laws. At the same time, some St\"ackel metrics
admit complete separation of variables even if they have no symmetry at all.

The St\"ackel problem was solved for inverse metrics of arbitrary signature with
nonzero diagonal elements long ago \cite{Acqua08,Burgat11,Acqua12,Havas75A}.
A different technique was used in \cite{KalMil80,KalMil81} (see also
\cite{Benent91}) for obtaining separable metrics including the case when
diagonal inverse metric components include zeroes. However the full list of
separating action functions and conservation laws in Hamiltonian formulation was
not derived. In recent paper \cite{Katana23A}, we propose complete solution to
this problem using another technique which allows us not only derive separable
metrics but, in addition, the full set of separating action functions and
conservation laws in the Hamiltonian formulation. In the present paper, we list
all separable metrics in four dimensions.

%*******************************************************************************
We consider $n$-dimensional topologically trivial manifold (diffeomorphic to
Euclidean space $\MR^n$) covered by a global coordinate system $x^\al$,
$\al=1,\dotsc,n$. Let there be a metric of arbitrary signature with components
$g_{\al\bt}(x)$ and its inverse $g^{\al\bt}(x)$:
$g_{\al\bt}g^{\bt\g}=\dl_\al^\g$ (summation over repeated indices is
understood). Then the kinetic Lagrangian for a line $x^\al(\tau)$, $\tau\in\MR$,
\begin{equation*}
  L_0:=\frac12g_{\al\bt}\dot x^\al\dot x^\bt,
\end{equation*}
where $\dot x^\al:=dx^\al/d\tau\ne0$ is the velocity of a line, produces
equations for geodesics parameterized by canonical parameter $\tau$. The
corresponding Hamiltonian is
\begin{equation}                                                  \label{anvbjd}
  H_0:=\frac12g^{\al\bt}p_\al p_\bt,
\end{equation}
where $p_\al:=\pl L_0/\pl\dot x^\al$ are momenta, yields Hamiltonian equations
for geodesics on a (pseudo)Riemannian manifold (see, e.g., \cite{DuNoFo98E}).

If metric is positive definite then Hamiltonian (\ref{anvbjd}) describes motion
of a point particle on Riemannian manifold. For Lorentzian signature metric,
Hamiltonian (\ref{anvbjd}) describes worldlines of point particles. Both cases
are of considerable interest from mathematical and physical points of view. The
Hamilton--Jacobi equation for the truncated action function $W(x)$ is
\begin{equation}                                                  \label{ehdggt}
  \frac12g^{\al\bt}\pl_\al W\pl_\bt W=E,\qquad E=\text{const}.
\end{equation}

A solution of Hamilton--Jacobi equation (\ref{ehdggt}) $W(x,c)$ depending on $n$
inde\-pendent parameters $c_a$, $a=1,\dotsc,n$, such that
\begin{equation}                                                  \label{indbfd}
  \det\frac{\pl^2 W}{\pl x^\al\pl c_a}\ne0,
\end{equation}
is called {\em complete integral}.

Coordinates $x^\al$, if they exist, are called {\em separable}, if
Hamilton--Jacobi equation (\ref{ehdggt}) admits additive separation of
variables in this coordinate system, i.e.\ the complete action function is given
by the sum
\begin{equation}                                                  \label{abcvde}
  W=\sum_{\al=1}^n W_\al(x^\al,c)
\end{equation}
where every summand $W_\al$ is a function of single coordinate $x^\al$ and,
possibly, a full set of $n$ parameters $c_a$.

For brevity, we use notation $W'_\al(x^\al,c):=\pl_\al W_\al(x^\al,c)$.
Then requirement (\ref{indbfd}) is
\begin{equation}                                                  \label{ijdhgf}
  \det(\pl W'_\al/\pl c_a)\ne0.
\end{equation}

Separable coordinates are not uniquely defined, and are divided into equivalence
classes. Two separable coordinate systems $x$ and $X$
are {\em equivalent}, if there is a canonical transformation $(x,p)\mapsto(X,P)$
of respective Hamiltonian systems, such that new coordinates $X(x)$ depend only
on old ones but not on momenta $p$.

For Lorentzian signature metrics, some coordinate lines $x^\al$ may be isotropic
when the respective metric component equals identically zero,
$g_{\al\al}(x)\equiv0$ (the tangent vector to this line is null). Separation of
variables depends essentially whether the inverse metric component is zero or
not. We call the coordinate line $x^\al$ {\em coisotropic}, when the
corresponding {\em inverse} metric component is zero, $g^{\al\al}(x)\equiv0$,
for brevity.

The number of independent conservation laws in the respective Hamiltonian system
is equal to the number of separating action functions $W'_\al$, and these
conservation laws are
in envolution (see, e.g.\ \cite{Katana23A}). For separable geodesic
Hamiltonians, there are three types of conservation laws: (i) there are linear
in momenta conservations laws related to Killing vector fields admitted by
separable metrics, (ii) conservation laws may be indecomposable quadratic (not
represented as a combination of linear conservation laws), and (iii) there are
linear conservation laws corresponding to coisotropic coordinates, for which
$g^{\al\al}\equiv0$. The last possibility arises only for indefinite metrics.

Suppose that there are exactly $\Sn\ge0$ and not more commuting Killing vectors,
$\Sm\ge0$ indecomposable quadratic conservation laws, and $n-\Sn-\Sm\ge0$
conservation laws corresponding to coisotropic coordinates. We divide all
coordinates onto three groups $(x^\al,y^\mu,z^\vf)$ where indices from the
beginning, middle, and end of the Greek alphabet take the following values:
\begin{equation}                                                  \label{abfgdr}
 \begin{aligned}
  \al,\bt,\dotsc=&1,\dotsc,\Sn & & \text{(commuting Killing vectors)},
\\
  \mu,\nu,\dotsc=&\Sn+1,\dotsc,\Sn+\Sm & & (\text{quadratic conservation laws},
  ~g^{\mu\mu}\ne0),
\\
 \vf,\phi,\dotsc=&\Sn+\Sm+1,\dotsc,n & & (\text{coisotropic coordinates},
 ~g^{\vf\phi}\equiv0).
\end{aligned}
\end{equation}
It means that separating action functions in Eq.(\ref{abcvde}) for Killing
vectors, quadratic conservation laws, and coisotropic coordinates depend on
single coordinates: $W_\al(x^\al,\dots)$, $W_\mu(y^\mu,\dots)$, and
$W_\vf(z^\vf,\dots)$, respectively, where dots denote parameters.

The full set of parameters is also divided onto three groups enumerated by
Latin letters from the beginning, middle, and end of the alphabet:
\begin{equation}                                                  \label{anvbfg}
  (c_a,d_{ij},a_r),\qquad a=1,\dotsc,\Sn,\quad
  i,j=\Sn+1,\dotsc,\Sn+\Sm,\quad r=\Sn+\Sm+1,\dotsc,n.
\end{equation}
where $d_{ij}$ is the diagonal matrix (to preserve the number of independent
parameters). We assume that energy $E$ enters the list
of independent parameters either as $a_n=2E$ (case 1), or
$d_{\Sn+\Sm\,\Sn+\Sm}=2E$ (case 2). Every separable metric belongs to one of the
following classes $[\Sn,\Sm,n-\Sn-\Sm]_{1,2}$, where the subscript denotes where
$E$ is located. All Riemannian separable metrics belong to classes
$[\Sn,\Sm,0]_2$, where $\Sn+\Sm=n$.

There are only 10 independent types of separable metrics in four dimensions. We
list the most simple (canonical) separable metrics in each class. All other
separable metrics are related to canonical metrics by suitable equivalence
transformations. Canonical metrics contain some arbitrary functions. We assume
that these functions produce nondegenerate metrics and expressions for $W'_\al$
are solvable for given parameters.

1) {\bf Type $[4,0,0]$.} Coordinates and parameters:
\begin{equation*}
  (x^\al,y^\mu,z^\vf)\mapsto(x^1,x^2,x^3,x^4),\qquad
  (c_\al,d_{ij},a_r)\mapsto(c_1,c_2,c_3,c_4).
\end{equation*}
The canonical separable metric is (pseudo)Euclidean:
\begin{equation*}
  g^{\al\bt}=\eta^{\al\bt},\qquad \al,\bt=1,2,3,4,
\end{equation*}
where $\eta^{\al\bt}$ is the diagonal matrix with $\pm1$ on the diagonal
depending on the signature of the metric. The
Hamilton--Jacobi equation is
\begin{equation*}
  \eta^{\al\bt}W'_\al W'_\bt=\eta^{\al\bt}c_\al c_\bt.
\end{equation*}
Variables are separated by $ W'_\al=c_\al$, and the Hamiltonian system has four
conservation laws $p_\al=c_\al$. There are four Killing vector fields $\pl_\al$.
The curvature tensor in this case vanishes and corresponding Cartesian
coordinates $x^\al$ are cyclic.

2) {\bf Type $[3,1,0]_2$.} Coordinates and parameters:
\begin{equation*}
  (x^\al,y^\mu,z^\vf)\mapsto(x^1,x^2,x^3,y),\qquad
  (c_\al,d_{ij},a_r)\mapsto(c_1,c_2,c_3,d_{44}:=2E).
\end{equation*}
The canonical separable metric is
\begin{equation}                                                  \label{abcvdy}
  g^{**}=\begin{pmatrix} g^{\al\bt}(y) & 0 \\ 0 & 1\end{pmatrix},
  \qquad \al,\bt=1,2,3,
\end{equation}
where $g^{\al\bt}$ is arbitrary symmetric nondegenerate matrix.
The Hamilton--Jacobi equation is
\begin{equation*}
  g^{\al\bt}W'_\al W'_\bt+W^{\prime2}_4=2E.
\end{equation*}
The variables are separated by
\begin{equation*}
  W'_\al=c_\al,\qquad\quad W^{\prime2}_4=2E-g^{\al\bt}c_\al c_\bt.
\end{equation*}
Conservation laws:
\begin{equation}                                                  \label{edjhyt}
  p_\al=c_\al,\qquad g^{\al\bt}p_\al p_\bt+p_4^2=2E.
\end{equation}
Depending on matrix $g^{\al\bt}$, the separable metric may have arbitrary
signature.

In what follows, we assume that all cyclic coordinates are separated,
$W'_\al\rightarrow c_\al$, and write equations only for the remaining
coordinates.

3) {\bf Type $[2,2,0]_2$.} Coordinates and parameters:
\begin{equation*}
  (x^\al,y^\mu,z^\vf)\mapsto(x^1,x^2,y^3,y^4),\qquad
  (c_\al,d_{ij},a_r)\mapsto(c_1,c_2,d_{33}:=d,d_{44}:=2E).
\end{equation*}
The canonical separable metric is
\begin{equation}                                                  \label{ancbfg}
  g^{**}=\frac1{\phi_3+\phi_4}\begin{pmatrix}
  k^{\al\bt}_3+k^{\al\bt}_4 & 0 & 0 \\ 0 & 1 & 0 \\ 0 & 0 & 1 \end{pmatrix}.
\end{equation}
where $\phi_3(y^3)$, $\phi_4(y^4)$, and $k^{\al\bt}_3(y^3)=k^{\bt\al}_3(y^3)$,
$k^{\al\bt}_4(y^4)=k^{\bt\al}_4(y^4)$ are arbitrary functions of indicated
arguments. The Hamilton--Jacobi equation becomes
\begin{equation*}
  \frac1{\phi_3+\phi_4}\big(k^{\al\bt}_3c_\al c_\bt
  +k^{\al\bt}_4c_\al c_\bt+W^{\prime2}_3+W^{\prime2}_4\big)=2E.
\end{equation*}
Variables are separated by
\begin{equation*}
\begin{split}
  W^{\prime2}_3=&~~d+2E\phi_3-k^{\al\bt}_3c_\al c_\bt,
\\
  W^{\prime2}_4=&-d+2E\phi_4-k^{\al\bt}_4c_\al c_\bt.
\end{split}
\end{equation*}
Conservation laws:
\begin{equation*}
\begin{split}
  \frac1{\phi_3+\phi_4}\big[\phi_4(p_3^2+k_3^{\al\bt}c_\al c_\bt)
  -\phi_3(p_4^2+k_4^{\al\bt}c_\al c_\bt\big]=&d,
\\
  \frac1{\phi_3+\phi_4}\big[p_3^2+k_3^{\al\bt}c_\al c_\bt+p_4^2+k^{\al\bt}_4
  c_\al c_\bt\big]=&2E.
\end{split}
\end{equation*}

Metric may have any signature depending on arbitrary functions.

4) {\bf Type $[2,0,2]_1$.} Coordinates and parameters:
\begin{equation*}
  (x^\al,z^\vf)=(x^1,x^2,z^3,z^4),\qquad
  (c_\al,d_{ii},a_r)\mapsto(c_1,c_2,a_3:=a,a_4:=2E).
\end{equation*}
The canonical separable metric is
\begin{equation}                                                  \label{aismnd}
  g^{**}=\frac1{2(1-\phi_3\phi_4)}\begin{pmatrix} 0& 0 & -\phi_4h_3^1 & h_4^1 \\
  0 & 0 & -\phi_4h_3^2 & h_4^2 \\ -\phi_4h_3^1 & -\phi_4h_3^2 & 0 & 0 \\
  h_4^1 & h_4^2 & 0 & 0 \end{pmatrix}.
\end{equation}
where $\phi_3(z^3)$, $\phi_4(z^4)$, $h_3^\al(z^3)$, and $h_4^\al(z^4)$,
$\al=1,2$, are arbitrary functions of single coordinates.
The Hamilton--Jacobi equation becomes
\begin{equation*}
  \frac1{1-\phi_3\phi_4}\big(-\phi_3h_3^\al c_\al W'_3+h_4^\al c_\al W'_4\big)
  =2E,
\end{equation*}
Variables are separated by
\begin{equation*}
\begin{split}
  W'_3=&\frac1{h_3^\al c_\al}(a+2\phi_3E),
\\
  W'_4=&\frac1{h_4^\al c_\al}(\phi_4a+2E).
\end{split}
\end{equation*}
Conservation laws:
\begin{equation*}
\begin{split}
  \frac1{1-\phi_3\phi_4}\big(h_3^\al c_\al p_3-\phi_3h_4^\al c_\al p_4\big)=&a,
\\
  \frac1{1-\phi_3\phi_4}\big(-\phi_4h_3^\al c_\al p_3+h^\al_4c_\al p_4\big)=&2E.
\end{split}
\end{equation*}
One of nonzero functions from each set $k_3^\al$ and $h_4^\al$ can be
transformed to unity by suitable canonical transformation. The canonical
separable metric (\ref{aismnd}) is parameterized by 4 arbitrary functions. Note
that $\det g^{**}$ is always positive. Therefore separable metrics of type
$[2,0,2]_1$ exist only for signature $(++--)$.

5) {\bf Type $[2,1,1]_1$.} Coordinates and parameters:
\begin{equation*}
  (x^\al,y^\mu,z^\vf)=(x^1,x^2,y,z),\qquad (c_\al,d_{ii},a_r)\mapsto
  (c_1,c_2,d_{33}:=d,a_4:=2E).
\end{equation*}
The canonical separable metric is
\begin{equation}                                                  \label{iekdwa}
  g^{**}=\frac1{1-\phi_3\phi_4} \begin{pmatrix}
  \phi_4k^{\al\bt} & 0 & h_4^\al/2 \\ 0 & -\phi_4 & 0 \\ h_4^\bt/2 & 0 & 0
  \end{pmatrix},
\end{equation}
where $\phi_3(y)$, $\phi_4(z)$, $h_4^\al(z)$, and $k^{\al\bt}(y)=k^{\bt\al}(y)$
are arbitrary functions of single coordinates.
The Hamilton--Jacobi equation becomes
\begin{equation*}
  \frac1{1-\phi_3\phi_4}\big(\phi_4k^{\al\bt}c_\al c_\bt-\phi_4W^{\prime2}_3
  +h_4^\al c_\al W'_4\big)=2E.
\end{equation*}
Variables are separated by
\begin{equation*}
\begin{split}
  W^{\prime2}_3=&d+2\phi_3E+k^{\al\bt}c_\al c_\bt,
\\
  W'_4=&\frac1{h^\al_4c_\al}\big(\phi_4d+2E\big).
\end{split}
\end{equation*}
Conservation laws:
\begin{equation*}
\begin{split}
  \frac1{1-\phi_3\phi_4}\big(p_3^2-k^{\al\bt}c_\al c_\bt
  -\phi_3h_4^\al c_\al p_4\big)=&d,
\\
  \frac1{1-\phi_3\phi_4}\big[-\ph_4(p_3^2-k^{\al\bt}c_\al c_\bt)
  +h_4^\al c_\al p_4\big]=&2E.
\end{split}
\end{equation*}
Making additional canonical transformation, we can put one of nonzero components
of $h_4^\al(z)$ to unity. Metric may have arbitrary signature depending on
arbitrary functions.

6) {\bf Type $[2,1,1]_2$.} Coordinates and parameters:
\begin{equation*}
  (x^\al,y^\mu,z^\vf)=(x^1,x^2,y,z),\qquad (c_\al,d_{ii},a_r)\mapsto
  (c_1,c_2,d_{33}:=2E,a_4=a),
\end{equation*}
The canonical separable metric is
\begin{equation}                                                  \label{ibjvdf}
  g^{**}=\frac1{1-\phi_3\phi_4}\begin{pmatrix}
  \displaystyle-k^{\al\bt} & 0 & -\phi_3h_4^\al/2 \\
  0 & 1 & 0 \\ -\phi_3h^\bt_4/2 & 0 & 0 \end{pmatrix},
\end{equation}
where $\phi_3(y)$, $\phi_4(z)$, $h_4^\al(z)$, and $k^{\al\bt}(y)=k^{\bt\al}(y)$,
$\al,\bt=1,2$, are arbitrary functions of single coordinates.
The Hamilton--Jacobi equation becomes
\begin{equation*}
  \frac1{1-\phi_3\phi_4}\big(-k^{\al\bt}c_\al c_\bt+W^{\prime2}_3
  -\phi_3h_4^\al c_\al W'_4\big)=2E.
\end{equation*}
Variables are separated by
\begin{equation*}
\begin{split}
  W^{\prime2}_3=&2E+\phi_3a+k^{\al\bt}c_\al c_\bt,
\\
  W'_4=&\frac1{h^\al_4c_\al}\big(2\phi_4E+a\big).
\end{split}
\end{equation*}
Conservation laws:
\begin{equation*}
\begin{split}
  \frac1{1-\phi_3\phi_4}\big(p_3^2-k^{\al\bt}c_\al c_\bt
  -\phi_3h_4^\al c_\al p_4\big)=&2E,
\\
  \frac1{1-\phi_3\phi_4}\big[-\phi_4(p_3^2-k^{\al\bt}c_\al c_\bt)
  +h^\al_4c_\al p_4\big]=&a.
\end{split}
\end{equation*}
Making additional canonical transformation, we can put one of nonzero components
of $h_4^\al$ to unity.

7) {\bf Type $[1,3,0]_2$.} Coordinates and parameters:
\begin{equation*}
  (x^\al,y^\mu,z^\vf)\mapsto(x,y^2,y^3,y^4),\qquad
  (c_\al,d_{ii},a_r)\mapsto(c,d_{22}:=d_2,d_{33}:=d_3,d_{44}:=2E).
\end{equation*}
Let nondegenerate matrix $b$ be
\begin{equation}                                                  \label{idukih}
  b_{\mu\mu}{}^{ii}=
  \begin{pmatrix}1 & b_{23}(y^2) & b_{24}(y^2)\\ b_{32}(y^3) & 1 & b_{34}(y^3)\\
  b_{42}(y^4) & b_{43}(y^4) & 1\end{pmatrix},\qquad
  b_{ii}{}^{\mu\mu}=\frac1\vartriangle
  \begin{pmatrix} \vartriangle_{22} & \vartriangle_{32} & \vartriangle_{42} \\
  \vartriangle_{23} & \vartriangle_{33} & \vartriangle_{43} \\
  \vartriangle_{24} & \vartriangle_{34} & \vartriangle_{44} \end{pmatrix},
\end{equation}
where elements are enumerated by pairs of indices $\mu\mu$ and $ii$ and entries
are arbitrary functions of single indicated coordinates,
$\vartriangle:=\det b_{\mu\mu}{}^{ii}$, and symbols $\vartriangle_{\mu i}$
denote cofactors of elements $b_{\mu\mu}{}^{ii}$. Then the canonical separable
metric is diagonal
\begin{equation}                                                  \label{ighsii}
  g^{**}=\frac1{\vartriangle}
  \begin{pmatrix}-\vartriangle_{24}k_2-\vartriangle_{34}k_3
    -\vartriangle_{44}k_4 & 0 & 0 & 0 \\
    0 & \vartriangle_{24} & 0 & 0 \\ 0 & 0 & \vartriangle_{34} & 0 \\
    0 & 0 & 0 & \vartriangle_{44} \end{pmatrix},
\end{equation}
where $k_2(y^2)$, $k_3(y^3)$, and $k_4(y^4)$ are arbitrary functions of single
coordinates. The Hamilton--Jacobi equation becomes
\begin{equation*}
  \frac1{\vartriangle}\big[\big(-\vartriangle_{24}k_2
  -\vartriangle_{34}k_3-\vartriangle_{44}k_4\big)c^2
  +\vartriangle_{24}W^{\prime2}_2+\vartriangle_{34}W^{\prime2}_3
  +\vartriangle_{44}W^{\prime2}_4\big]=2E.
\end{equation*}
Variables are separated by
\begin{equation*}
\begin{split}
  W^{\prime2}_2=d_2+b_{23}d_3+2b_{24}E+k_2c^2,
\\
  W^{\prime2}_3=b_{32}d_2+d_3+2b_{34}E+k_3c^2,
\\
  W^{\prime2}_4=b_{42}d_2+b_{43}d_3+2E+k_4c^2.
\end{split}
\end{equation*}
Conservation laws:
\begin{equation*}
\begin{split}
  \frac1\vartriangle\big[\vartriangle_{22}(p_2^2-k_2c^2)
  +\vartriangle_{32}(p_3^2-k_3c^2)+\vartriangle_{42}(p_4^2-k_4c^2)\big]=&d_2,
\\
  \frac1\vartriangle\big[\vartriangle_{23}(p_2^2-k_2c^2)
  +\vartriangle_{33}(p_3^2-k_3c^2)+\vartriangle_{43}(p_4^2-k_4c^2)\big]=&d_3,
\\
  \frac1\vartriangle\big[\vartriangle_{24}(p_2^2-k_2c^2)
  +\vartriangle_{34}(p_3^2-k_3c^2)+\vartriangle_{44}(p_4^2-k_4c^2)\big]=&2E.
\end{split}
\end{equation*}
Metric may have arbitrary signature depending on arbitrary functions.

8) {\bf Type $[1,2,1]_2$.} Coordinates and parameters:
\begin{equation*}
  (x^\al,y^\mu,z^\vf)\mapsto(x,y^2,y^3,z),\qquad
  (c_\al,d_{ij},a_r)\mapsto(c,d_{22}:=d,d_{33}:=2E,a).
\end{equation*}
Let nondegenerate matrix $B$ be
\begin{equation}                                                  \label{idshfd}
  B=\begin{pmatrix} 1 & \phi_{23} & \phi_{24} \\ \phi_{32} & 1 & \phi_{34} \\
  \phi_{42}/c & \phi_{43}/c & 1/c \end{pmatrix}\qquad\Rightarrow\qquad
  B^{-1}=\frac1\vartriangle
  \begin{pmatrix} \vartriangle_{22} & \vartriangle_{32} & \vartriangle_{42} \\
  \vartriangle_{23} & \vartriangle_{33} & \vartriangle_{43} \\
  \vartriangle_{24} & \vartriangle_{34} & \vartriangle_{44} \end{pmatrix},
\end{equation}
where $\phi_{23}(y^2)$, $\phi_{24}(y^2)$, $\phi_{32}(y^3)$, $\phi_{34}(y^3)$,
$\phi_{42}(z)$, and $\phi_{43}(z)$ are arbitrary functions of single
coordinates. In addition,
\begin{equation*}
  \vartriangle:=1+\phi_{23}\phi_{34}\phi_{42}+\phi_{24}\phi_{43}\phi_{32}
  -\phi_{23}\phi_{32}-\phi_{24}\phi_{42}-\phi_{34}\phi_{43}
\end{equation*}
and
\begin{equation*}
\begin{aligned}
  \vartriangle_{22}=& 1-\phi_{34}\phi_{43}, &
  \qquad\vartriangle_{32}=& \phi_{24}\phi_{43}-\phi_{23}, &
  \qquad\vartriangle_{42}=& (\phi_{23}\phi_{34}-\phi_{24})c,
\\
  \vartriangle_{23}=& \phi_{34}\phi_{42}-\phi_{32}, &
  \vartriangle_{33}=& 1-\phi_{24}\phi_{42}, &
  \vartriangle_{43}=& (\phi_{32}\phi_{24}-\phi_{34})c,
\\
  \vartriangle_{24}=& \phi_{43}\phi_{32}-\phi_{42}, &
  \vartriangle_{34}=& \phi_{42}\phi_{23}-\phi_{43}, &
  \vartriangle_{44}=& (1-\phi_{23}\phi_{32})c.
\end{aligned}
\end{equation*}
The canonical separable metric is
\begin{equation}                                                  \label{idkwey}
  g^{**}=\frac1\vartriangle\begin{pmatrix}
  -\vartriangle_{23}k_2-\vartriangle_{33}k_3 & 0 & 0 & \vartriangle_{43}/(2c)\\
  0 & \vartriangle_{23} & 0 & 0 \\ 0 & 0 & \vartriangle_{33} & 0 \\
  \vartriangle_{43}/(2c) & 0 & 0 & 0 \end{pmatrix},
\end{equation}
where $k_2(y^2)$ and $k_3(y^3)$ are arbitrary functions.
The Hamilton--Jacobi equation becomes
\begin{equation*}
  \frac1\vartriangle\left[-(\vartriangle_{23}k_2+\vartriangle_{33}k_3)c^2
  +\vartriangle_{23}W^{\prime2}_2+\vartriangle_{33}W^{\prime2}_3
  +\vartriangle_{43}W'_4\right]=2E.
\end{equation*}
Variables are separated by
\begin{equation*}
\begin{split}
  W^{\prime2}_2=&d+2\phi_{23}E+\phi_{24}a+k_2c^2,
\\
  W^{\prime2}_3=&\phi_{32}d+2E+\phi_{34}a+k_3c^2,
\\
  W'_4=&\frac1c\big(\phi_{42}d+2\phi_{43}E+a\big).
\end{split}
\end{equation*}
Conservation laws:
\begin{equation*}
\begin{split}
  \frac1\vartriangle\big[\vartriangle_{22}(p_2^2-k_2c^2)
  +\vartriangle_{32}(p_3^2-k_3c^2)+\vartriangle_{42}p_4\big]=&d,
\\
  \frac1\vartriangle\big[\vartriangle_{23}(p_2^2-k_2c^2)
  +\vartriangle_{33}(p_3^2-k_3c^2)+\vartriangle_{43}p_4\big]=&2E,
\\
  \frac1\vartriangle\big[\vartriangle_{24}(p_2^2-k_2c^2)
  +\vartriangle_{34}(p_3^2-k_3c^2)+\vartriangle_{44}p_4\big]=&a,
\end{split}
\end{equation*}
Metric may have arbitrary signature depending on arbitrary functions.

9) {\bf Type $[1,2,1]_1$.} Coordinates and matrix $B$ are the same
as for separable metrics of type $[1,2,1]_2$. Parameters:
\begin{equation*}
  (c_\al,d_{ij},a_r)\mapsto(c,d_{22}:=d_2,d_{33}:=d_3,a_4:=2E).
\end{equation*}
The canonical separable metric is
\begin{equation}                                                  \label{idkwel}
  g^{**}=\frac1\vartriangle\begin{pmatrix}
  -\vartriangle_{24}k_2-\vartriangle_{34}k_3 & 0 & 0 & \vartriangle_{44}/(2c)\\
  0 & \vartriangle_{24} & 0 & 0 \\ 0 & 0 & \vartriangle_{34} & 0 \\
  \vartriangle_{44}/(2c) & 0 & 0 & 0 \end{pmatrix}.
\end{equation}
The Hamilton--Jacobi equation becomes
\begin{equation*}
  \frac1\vartriangle\left[-(\vartriangle_{24}k_2+\vartriangle_{34}k_3)c^2
  +\vartriangle_{24}W^{\prime2}_2+\vartriangle_{34}W^{\prime2}_3
  +\vartriangle_{44}W'_4\right]=2E.
\end{equation*}
Variables are separated by
\begin{equation*}
\begin{split}
  W^{\prime2}_2=&d_2+\phi_{23}d_3+2\phi_{24}E+k_2c^2,
\\
  W^{\prime2}_3=&\phi_{32}d_2+d_3+2\phi_{34}E+k_3c^2,
\\
  W'_4=&\frac1c\big(\phi_{42}d_2+\phi_{43}d_3+2E\big).
\end{split}
\end{equation*}
Conservation laws:
\begin{equation*}
\begin{split}
  \frac1\vartriangle\big[\vartriangle_{22}(p_2^2-k_2c^2)
  +\vartriangle_{32}(p_3^2-k_3c^2)+\vartriangle_{42}p_4\big]=&d_2,
\\
  \frac1\vartriangle\big[\vartriangle_{23}(p_2^2-k_2c^2)
  +\vartriangle_{33}(p_3^2-k_3c^2)+\vartriangle_{43}p_4\big]=&d_3,
\\
  \frac1\vartriangle\big[\vartriangle_{24}(p_2^2-k_2c^2)
  +\vartriangle_{34}(p_3^2-k_3c^2)+\vartriangle_{44}p_4\big]=&2E.
\end{split}
\end{equation*}
Metric may have arbitrary signature depending on arbitrary functions.

10) {\bf Type $[0,4,0]_2$.} Coordinates and parameters:
\begin{equation*}
  (x^\al,y^\mu,z^\vf)\mapsto(y^1,y^2,y^3,y^4),\qquad
  (c_\al,d_{ij},a_r)\mapsto(d_1,d_2,d_3,d_{44}:=2E).
\end{equation*}
Let nondegenerate matrix $B$ be
\begin{equation}                                                  \label{idlkih}
  b_{\mu\mu}{}^{ii}=
  \begin{pmatrix}1 & b_{12}(y^1) & b_{13}(y^1) & b_{14}(y^1) \\
  b_{21}(y^2) & 1 & b_{23}(y^2) & b_{24}(y^2) \\
  b_{31}(y^3) & b_{32}(y^3) & 1 & b_{34}(y^3) \\
  b_{41}(y^4) & b_{42}(y^4) & b_{43}(y^4) & 1\end{pmatrix},\quad
  b_{ii}{}^{\mu\mu}=\frac1\vartriangle
  \begin{pmatrix}
  \vartriangle_{11} & \vartriangle_{21} & \vartriangle_{31} &\vartriangle_{41}\\
  \vartriangle_{12} & \vartriangle_{22} & \vartriangle_{32} &\vartriangle_{42}\\
  \vartriangle_{13} & \vartriangle_{23} & \vartriangle_{33} &\vartriangle_{43}\\
  \vartriangle_{14} & \vartriangle_{24} & \vartriangle_{34} &\vartriangle_{44}\\
  \end{pmatrix},
\end{equation}
where $\vartriangle:=\det b_{\mu\mu}{}^{ii}$, and symbols
$\vartriangle_{\mu i}$ denote cofactors of elements $b_{\mu\mu}{}^{ii}$. The
canonical separable metric is
\begin{equation}                                                  \label{avcfsg}
  g^{**}=\frac1\vartriangle \begin{pmatrix} \vartriangle_{14} & 0 & 0 & 0 \\
  0 & \vartriangle_{24} & 0 & 0 \\ 0 & 0 & \vartriangle_{34} & 0 \\
  0 & 0 & 0 & \vartriangle_{44} \end{pmatrix}.
\end{equation}
The Hamilton--Jacobi becomes
\begin{equation*}
  \frac1\vartriangle\big(\vartriangle_{14}W^{\prime2}_1
  +\vartriangle_{24}W^{\prime2}_2+\vartriangle_{34}W^{\prime2}_3
  +\vartriangle_{44}W^{\prime2}_4\big)=2E.
\end{equation*}
Variables are separated by
\begin{equation*}
\begin{split}
  W^{\prime2}_1=&d_1+b_{12}d_2+b_{13}d_3+2b_{14}E,
\\
  W^{\prime2}_2=&b_{21}d_1+d_2+b_{23}d_3+2b_{24}E,
\\
  W^{\prime2}_3=&b_{31}d_1+b_{32}d_2+d_3+2b_{34}E.
\\
  W^{\prime2}_4=&b_{41}d_1+b_{42}d_2+b_{43}d_3+2E.
\end{split}
\end{equation*}
Conservation laws:
\begin{equation*}
\begin{split}
  \frac1\vartriangle\big(\vartriangle_{11}p^2_1+\vartriangle_{21}p^2_2
  +\vartriangle_{31}p^2_3+\vartriangle_{41}p^2_4\big)=&d_1,
\\
  \frac1\vartriangle\big(\vartriangle_{12}p^2_1+\vartriangle_{22}p^2_2
  +\vartriangle_{32}p^2_3+\vartriangle_{42}p^2_4\big)=&d_2,
\\
  \frac1\vartriangle\big(\vartriangle_{13}p^2_1+\vartriangle_{23}p^2_2
  +\vartriangle_{33}p^2_3+\vartriangle_{43}p^2_4\big)=&d_3.
\\
  \frac1\vartriangle\big(\vartriangle_{14}p^2_1+\vartriangle_{24}p^2_2
  +\vartriangle_{34}p^2_3+\vartriangle_{44}p^2_4\big)=&2E.
\end{split}
\end{equation*}

It is easily checked that all metrics of the present paper are really separable.
The proof that this is the complete list of separable metrics is elementary but
lengthy. The details are given in \cite{Katana23A}.

Separable metrics in four dimensions were listed in \cite{BoKaMi78}. The types
of metrics are the same, but we give more detailed classification denoted by
indices $[\Sn,\Sm,n-\Sn-\Sm]_{1,2}$. We also succeeded in finding explicit
expressions for all separating functions $W'_\al$, $\al=1,\dotsc,n$ and all
conservation laws. Seven of ten separable metrics are listed in \cite{Obukho06}.

We see that separable metrics depend on many arbitrary functions. The
corresponding curvature tensor differs from zero in general. If the curvature is
identically zero, then separable metrics are written in curvilinear coordinates
on a flat manifolds. For nontrivial curvature, we have separable metrics on a
(pseudo)Riemannian manifolds. In particular, the manifold may be of constant
curvature (this depends on arbitrary functions defining separable metrics).

%*******************************************************************************
Separable metrics of Lorentzian metrics are of great importance in general
relativity.
B.~Carter analyzed Einstein--Maxwell equations with cosmological constant for
separable metrics of type $[2,2,0]_2$ and showed that Schwarzschild,
Reissner--Nordstr\"om, Kerr and many more solutions belong to this class
\cite{Carter68}. It is easily checked that Kasner solution is of type
$[3,1,0]_2$ and Friedmann solution is of type $[1,3,0]_2$. So there is an
interesting problem to check to which class belongs a known exact solution. It
is nontrivial because solutions are often written in a nonseparable coordinate
systems. The opposite way of thought is to analyze Einstein equations for all
other separable metrics like it was done by B.~Carter. Moreover, geodesic
equations in such cases are Liouville integrable, and the global structure of
respective space-times can be, probably, analyzed analytically.

%\bibliography{2dgrav,3dgrav,book,gravity,math,my,qft}
%\bibliographystyle{unsrt}
\end{document}